\documentclass[prl,aps,showpacs]{revtex4}

\usepackage{verbatim}
\usepackage{epsfig}
\usepackage{amssymb}
\usepackage{amsmath}
\usepackage{bm}

\begin{document}

\title{The Signature of Proper Motion in the Microwave Sky}

\date{\today}
\author{Arthur Kosowsky}
\email{kosowsky@pitt.edu}
\affiliation{Department of Physics and
Astronomy, University of Pittsburgh, 3941 O'Hara Street, Pittsburgh, PA 15260 USA}
\author{Tina Kahniashvili}
\affiliation{Department of Physics, Carnegie Mellon University, 5000 Forbes Ave., Pittsburgh, PA 15213 USA}
 \affiliation{Department of Physics, Laurentian University, Ramsey Lake Road, Sudbury, ON P3E 2C6,
Canada} 
\affiliation{Abastumani Astrophysical Observatory, Ilia Chavchavadze State University, 2A Kazbegi Ave., Tbilisi, GE-0160, Georgia}

\begin{abstract}
The cosmic microwave background radiation defines a preferred cosmic rest frame, and
inflationary cosmological theories predict that the microwave background temperature fluctuations
should be statistically isotropic in this rest frame. For observers moving with respect to
the rest frame, the temperature fluctuations will no longer be isotropic, due to the preferred
direction of motion. The most prominent effect is a dipole temperature variation, which
has long been observed with an amplitude of a part in a thousand of the mean temperature. 
An observer's velocity with respect to the rest frame will also induce 
changes in the angular correlation function and
creation of non-zero off-diagonal correlations between multipole moments. We calculate
both of these effects, which are part-in-a-thousand corrections to the rest frame
power spectrum and correlation function. 
Both should be detectable in future full-sky microwave maps from the Planck satellite. 
These signals will constrain cosmological models in which the cosmic dipole arises partly from
large-scale isocurvature perturbations, as suggested by recent observations.
\end{abstract}


\maketitle
The most prominent feature in the microwave background radiation is the large dipole
modulation, at a part in a thousand of the mean background temperature
\cite{con69,hen71,cor76,smo77,kog93,ben03}. This is generally
attributed to our local peculiar velocity with respect to the cosmic rest frame of 370 km/s
\cite{bra68,hen68,pee68,wei72}.
However, changing frames from the cosmic rest frame to a boosted frame also induces
small, distinctive changes in both the cross-power spectrum and the correlation function
of the microwave radiation, both of which are potentially detectable in full sky maps with
the angular resolution of the Planck satellite. Here we present a straightforward calculation of the signals
and discuss their detectability, and note that subtle microwave background distortions
are a promising route for constraining ``tilted'' cosmological models where an isocurvature
perturbation on the scale of the horizon contributes to the microwave dipole and to
large-scale streaming motions of galaxies and galaxy clusters. 

Similar calculations were pioneered by Challinor and van Leeuwen \cite{cha01}, but they
were primarily concerned with demonstrating that the effects were small enough to be 
neglected when constraining cosmological parameters with the microwave background power spectrum;
they did not consider the correlation function, and did not consider detectability of the signals.
Burles and Rappaport \cite{bur06} considered detectability of the aberration in the microwave
radiation via the shift in angular scale of acoustic peaks it
introduces; this effect is related to the correlation function analysis we present here. 

\section{Lorentz Transformation of the Temperature Field}

A clear discussion of how a radiation field transforms under Lorentz boosts has been given by
Ref.~\cite{mck79}, which clarifies some misconceptions in earlier literature,
and the discussion here follows this paper. Consider 
a frame $S'$ which is the rest frame of the microwave background, so that $T'({\bf n}')$ is the
temperature distribution in this frame. Now take frame $S$ to be our observation frame which is
boosted from the cosmic rest frame by a velocity ${\bf v}$, with resulting sky temperature $T({\bf n})$
(we use velocity units with $c=1$). 
Theories of cosmology predict a sky map in the cosmic rest frame, while we observe
a sky map in our boosted frame; here we obtain the transformation of the correlations between
multipole moments and the transformation of the correlation function between the rest frame and
the boosted frame. 

Consider a photon with wavevector ${\bf k}'=k'{\bf n}'$ in the rest frame and wavevector ${\bf k}=k{\bf n}$ in the observation frame. 
We have the wavenumber transformation
\begin{equation}
k = \gamma(1+{\bf v}\cdot{\bf n}')k' \equiv Dk'
\label{ktransform}
\end{equation}
and $dV = D^{-1}dV'$ for the transformation of a volume element. 
We also have the aberration equation
\begin{equation}
{\bf\hat v}\cdot{\bf n} = \frac{{\bf\hat v}\cdot{\bf n}' + v}{1 + {\bf v}\cdot{\bf n}'}
\label{costheta_transf}
\end{equation}
with ${\bf\hat v}$ a unit vector in the velocity direction.
Taking derivatives of these two equations gives the transformation of the differential
solid angle
$d{\bf n}' \equiv \sin\theta' d\theta' d\phi$
as $d{\bf n} = D^{-2}d{\bf n}'$. 

In the rest frame $S'$, the number of photons in a volume $dV'$ which have a wavevector $k'{\bf n}'$ in the
interval $dk'\,d{\bf n}'$ is given by
\begin{equation}
dN = \frac{1}{4\pi^3}\frac{1}{\exp(\hbar k'/T'({\bf n}')) -1} k'^2 dk' dV' d{\bf n}' 
\label{dN_CMBrest}
\end{equation}
which is from the definition of a blackbody photon distribution. This same number of photons viewed in
frame $S$ will have propagation vector $k{\bf n}$ in the interval $dk\, d{\bf n}$ and 
will occupy a volume $dV$. 
Rewriting Eq.~(\ref{dN_CMBrest}) in terms of the transformed variables 
gives
\begin{equation}
dN = \frac{1}{4\pi^3}\frac{1}{\exp(\hbar k/DT'({\bf n}')) -1} k^2 dk\,dV d{\bf n} .
\label{dN_CMBobs}
\end{equation}
Note that the combination $k^2 dk\, dV d{\bf n}$ is invariant, so the only change comes in the argument of
the exponential. Now again using the blackbody definition, we identify 
\begin{equation}
T({\bf n}) = DT'({\bf n}').
\label{T_transform}
\end{equation} 
This gives the transformation between the CMB rest-frame temperature distribution $T'({\bf n}')$
and the observation-frame temperature distribution $T({\bf n})$. 
From this,
we now derive the transformation of observables which are commonly extracted from cosmological models,
namely correlations of multipole moments and the angular correlation function.

\section{Transformation of Power Spectra and the Correlation Function}

The (rest-frame) microwave
sky temperature is commonly expressed in terms of spherical harmonics,
\begin{equation}
T'({\bf n}') = \sum_{lm}a'_{lm}Y_{lm}({\bf n}'),
\label{alm_def}
\end{equation}
where the angular power spectrum in terms of these coefficients is
$C'_l = \langle a_{lm}^{\prime\,*} a'_{lm}\rangle$. Here the angle brackets refer to an ensemble average over realizations
of a random temperature field on the sky with the same underlying statistical properties. If the rest frame universe  is statistically isotropic, then each moment $C'_l$ of the angular power spectrum is independent of $m$, and
the average value of coefficients with different indices vanishes: $\langle a_{l'm'}^{\prime\,*} a'_{lm}\rangle = 0$ if $l\neq l'$
or $m\neq m'$. We want the transformation law connecting the coefficients in the two frames.

The individual $a_{lm}$ values transform as follows:
\begin{equation}
a_{lm} = \int d{\bf n}\, T({\bf n}) Y^*_{lm}({\bf n})
=\int d{\bf n} \frac{1+{\bf v}\cdot{\bf n}'}{\sqrt{1-v^2}} T'({\bf n}')Y_{lm}^*({\bf n}).
\end{equation}
Now we choose a spherical coordinate system with the $\bf z$-axis aligned with the boost direction, and
change integration variables to the rest-frame angles with ${\bf n}' = (\theta',\phi)$,
\begin{equation}
a_{lm} = \int_0^\pi \sin\theta' d\theta'\int_0^{2\pi} d\phi 
\frac{\sqrt{1-v^2}}{1+v\cos\theta'} T'(\theta',\phi)
Y^*_{lm}\left(\frac{\cos\theta'+v}{1+v\cos\theta'},\phi\right).
\end{equation}
Then expanding the rest-frame temperature distribution in spherical harmonics and
doing the trivial integral over $\phi$ gives the exact expression 
\begin{equation}
a_{lm} = \sum_{l'=0}^\infty a'_{l'm} I^m_{l'l}(v)
\label{alm_equation}
\end{equation}
(no sum over $m$) where we have defined
\begin{equation}
I^m_{l'l}(v)\equiv 2\pi\sqrt{1-v^2} \int_{-1}^1 \frac{dx}{1+v x} {\tilde P}^m_{l'}(x)
{\tilde P}^m_l\left(\frac{x+v}{1+v x}\right)
\label{I_definition}
\end{equation}
with the abbreviation
\begin{equation}
{\tilde P}_l^m(x) = \left(\frac{2l+1}{4\pi}\frac{(l-m)!}{(l+m)!}\right)^{1/2}P_l^m(x)
\label{Ptilde}
\end{equation}
for the spherical-harmonic-normalized associated Legendre functions. 

Some care must be taken in the numerical
evaluation of Eq.~(\ref{I_definition}), since the integrand is rapidly oscillating for
large values of $l-|m|$ and underflows in the integrand can compromise convergence
conditions with adaptive numerical integrators.
Direct numerical integration reveals the orthonormality relation
\begin{equation}
\sum_{l'} I^{m}_{l'l_1}I^{m}_{l'l_2} = \delta_{l_1 l_2}
\label{I_orthog}
\end{equation}
(no sum on $m$) which is valid for any velocity $v$. 
For small $v$, 
Eq.~(\ref{I_definition}) has the asymptotic behavior
\begin{eqnarray}
I_{ll}^m &\sim& 1 + {\cal O}(v^2),\qquad v\rightarrow 0 \label{Illmlin}\\
I_{l+1,l}^m &\sim& -v(l+1)\sqrt{\frac{(l+1)^2-m^2}{(2l+1)(2l+3)}},\qquad v\rightarrow 0 \\
I_{l-1,l}^m &\sim& v l\sqrt{\frac{l^2-m^2}{(2l+1)(2l-1)}},\qquad v\rightarrow 0 
\end{eqnarray}
with all other $l',l$ values being ${\cal O}(v^2)$ or smaller. 

From Eq.~(\ref{alm_equation}), we have the boosted-frame products of coefficients
\begin{equation}
\left\langle a^*_{l_1m_1}a^{\phantom{*}}_{l_2m_2}\right\rangle = \delta_{m_1m_2} \sum_{l'} C'_{l'} 
I^{m_1}_{l'l_1}I^{m_2}_{l'l_2}.
\label{Cl_transform}
\end{equation}
Note that the statistical ensemble averaging procedure on the left side of this expression is independent
of frame. When $l_2=l_1$, the linear perturbative expansion in  Eq.~(\ref{Illmlin}) is not sufficient: 
perturbative evaluation of $C_l$ in the boosted frame requires also the ${\cal O}(\beta^2)$ term for
consistency, and values of $l\gtrsim 1/v$ formally require increasingly higher powers of $v$ for
convergence. Using the rest-frame power spectrum
$C'_l$ given by the WMAP best-fit cosmology \cite{WMAP5}, 
direct numerical evaluation of Eq.~(\ref{Cl_transform}) with $v = 0.00123$ 
gives that $C_1\approx 5000 C'_1$ for
the observed dipole, and the fractional corrections
$(C'_2 - C_2)/C_2 \approx 6\times 10^{-3}$ for the quadrupole and $(C'_l - C_l)/C_l$ ranging
between $10^{-6}$
and $5\times 10^{-5}$ for all $3\leq l<1500$. These corrections to the power spectrum are
too small to be observed, given the cosmic variance.  Correlations with $|l_2 - l_1| \geq 2$
are at most ${\cal O}(v^2)$ or smaller and also undetectably small as verified by direct
numerical calculation.

However, for the case $l_2=l_1+1$, 
the linear expansions in $v$ for $I_{ll'}^m$ are sufficient for a  consistent evaluation,
yielding 
\begin{equation}
\left\langle a^*_{l+1,m} a^{\phantom{*}}_{lm}\right\rangle \simeq (C'_{l+1}-C'_l)v (l+1)
\sqrt{\frac{(l+1)^2-m^2}{(2l+3)(2l+1)}}
\label{alp1al}
\end{equation}
up to corrections of ${\cal O}(v^2)$. Since roughly $C'_l \approx l^{-2} C'_ 2$ for $l$ up to
roughly 1000 (neglecting acoustic oscillations), $C'_{l+1}-C'_l \approx -2C'_l/l$
and $\langle a^*_{l+1,m} a_{lm}\rangle\approx -v C'_l$ for large $l$
and small $m$. This signal can be detected statistically, as shown below.

The change in the two-point correlation function is also interesting, since
distortions in the shapes of microwave hot and cold spots due to the boost is
an effect in angle $\theta$ space and not in multipole $l$ space. For two sky directions
${\bf n}_1$ and ${\bf n}_2$, the two-point correlation function is defined as
\begin{equation}
C({\bf n}_1,{\bf n}_2) \equiv \left\langle T({\bf n}_1)T({\bf n}_2)\right\rangle .
\label{Cdef}
\end{equation}
In the rest frame, which we assume to be statistically isotropic, the correlation
function $C'({\bf n}_1^\prime, {\bf n}_2^\prime)$ depends only on the angle between the two observation directions
${\bf n}_1^\prime\cdot{\bf n}_2^\prime$. 
Substituting Eq.~(\ref{T_transform}) into Eq.~(\ref{Cdef}) yields
\begin{equation} 
C({\bf n}_1, {\bf n}_2) \simeq (1 + {\bf v}\cdot{\bf n}_1^\prime + {\bf v}\cdot{\bf n}_2^\prime)
C'({\bf n}_1^\prime \cdot {\bf n}_2^\prime)
\label{Cboostfinal}
\end{equation}
up to  ${\cal O}(v^2)$ corrections.
As statistical isotropy is broken by the Lorentz boost, the correlation function now depends on the
two directions separately. The rest-frame correlation function is modified at ${\cal O}(v)$ and
varies with the angle between the boost direction and the observation direction. 

\section{Detectability and Utility}

For a full-sky microwave temperature map with $N_{\rm pix}$ pixels, each with Gaussian
noise $\sigma_{\rm pix}$, and a Gaussian beam of width $\sigma_b$, each $a_{lm}$
is approximately normally distributed with a 
variance $C_l\exp(-l^2\sigma_b^2) + w^{-1}$ \cite{kno95} and uncorrelated with other
$a_{lm}$ values, where
$w^{-1}\equiv 4\pi \sigma_{\rm pix}^2/N_{\rm pix}$ is the inverse statistical weight per unit
solid angle. Solving Eq.~(\ref{alp1al})
for $v$, averaging over $l$ and $m$ with signal-to-noise weighting, 
and propagating errors from the $a_{lm}$ gives
a standard error $\sigma_v$ on $v$ from a full-sky map using multipoles 
$2\leq l  < l_{\rm max}$ of
\begin{equation}
\sigma_v = l_{\rm max}^2\left[\sum_{l=2}^{l_{\rm max}-1} \sum_{m=-l}^l 
(l+1)\left(\frac{(l+1)^2-m^2}{2(2l+1)(2l+3)}\right)^{1/2}
\left(1-\frac{C_{l+1}}{C_l}\right)
\left(e^{-l^2\sigma_b^2} + \frac{w^{-1}}{C_l}\right)^{-1/2}\right]^{-1}.
\label{verror}
\end{equation}
The Planck satellite's 143 GHz channel has approximately $\sigma_b = 3.1$ arcminutes,
and $N_{\rm pix} = 2.9\times 10^6$ with a target noise level of $\sigma_{\rm pix}= 6.0$ $\mu$K, giving 
$w^{-1}=1.6\times 10^{-4}$ $\mu{\rm K}^2$. $l_{\rm max}$ is determined by the largest $l$ for
which systematic errors in beam characterization do not dominate the error model for $a_{lm}$. For
$l_{\rm max}=2000$, Eq.~(\ref{verror}) gives $\sigma_v = 2.5\times 10^{-4}$. If the dipole is due entirely to our peculiar velocity,
$v=0.00123$ and Planck can detect this signal through the off-diagonal cross-power signal
at a signal-to-noise ratio of 5. For $l_{\rm max}=2500$, the signal-to-noise ratio increases to 6; other
Planck channels will provide independent estimates and further increase the signal-noise ratio.
Foreground emission and partial sky coverage may in practice reduce somewhat the significance of
a detection, although neither has greatly impacted measurements of the temperature power spectrum.

Detectability of the small corrections in Eq.~(\ref{Cboostfinal}) 
is harder to estimate, since
values of the correlation function for similar angles are highly
correlated. In general, a map contains more information about the correlation function than
about the power spectrum, since the correlation function is significant 
out to separations as large as $50^\circ$. At a separation $\theta$, a map has
approximately $(2\pi\theta/\sigma_b) N_{\rm pix}$ pairs; for the Planck map above and,
e.g., $\theta=10^\circ$, this is about $3.5\times 10^9$ pairs. Averaging over all pairs of
pixels, each with Gaussian error $\sigma_{\rm pix}$, and propagating through the statistical
errors on each pixel gives the standard error on $C(\theta)$ as 
$\sigma_\theta = \sigma_{\rm pix}\sqrt{2C(0)/N_{\rm pairs}}$.
For a monopole and dipole-subtracted map, $C(0)=\sum_l (2l+1)C_l\exp(-l^2\sigma_b^2)/(4\pi) =  1.1\times 10^4$ $\mu{\rm K}^2$ for the Planck beam above, so $\sigma_\theta =  0.015$ $\mu{\rm K}^2$,
compared to a signal of $C(\theta = 10^\circ)\gtrsim 1000$ $\mu{\rm K}^2$ \cite{cop07}. Testing the form
of Eq.~(\ref{Cboostfinal}) requires comparing different portions of the sky for a variation in the
correlation function of a part in a thousand. The correlation function can be estimated at many different angles, with each providing a moderate signal-to-noise measurement of the difference in the correlation
function between different sky regions. However, this estimate includes only instrumental noise,
and does not account for cosmic variance between regions; more precise detectability estimates require
evaluation of both the signal covariance for different angles and cosmic variance for different
regions (e.g., \cite{hin96}). Correlations from foreground emission are also a challenge for this
measurement. 

Aside from being a consistency check on a fundamental cosmological property,
the distinctive microwave background signals from a local velocity with respect to the
microwave background rest frame will constrain ``tilted'' cosmological models where the dipole
arises partly due to primordial superhorizon-scale isocurvature fluctuations 
\cite{tur91,gri92,lan96,eri08,mer09}.
Such models naturally explain surprising recent observations of a substantial galaxy cluster bulk flow on
Hubble volume scales \cite{kas10} and
galaxy bulk flow on somewhat smaller scales \cite{wat09}. 

\begin{acknowledgments}
We thank E.T.~Newman for prompting this analysis, and 
D.N.~Spergel for helpful comments. C.~Copi and G.~Starkmann have
also pursued this issue, and after this paper was submitted for publication we were made
aware of a similar paper in preparation by L.~Amendola, R.~Catena, I.~Masina, A.~Notari,
M.~Quartin, and C.~Quercellini.
This work made use of the GNU Scientific Library for numerical routines, and
the NASA Astrophysical Data System for bibliographic information. The authors
are supported by NASA Astrophysics Theory Program grant NNXlOAC85G.
TK is supported by Georgian National Science Foundation grant GNSF ST08/4-422
and Swiss National Science Foundation SCOPES grant 128040. AK is supported
by NSF grant 0807790.
\end{acknowledgments}



\begin{thebibliography}{}

\bibitem{con69}
E.K.~Conklin, Nature 222, 971 (1969). 

\bibitem{hen71}
P.S.~Henry, Nature 231, 516 (1971).

\bibitem{cor76}
B.E.~Corey and D.T.~Wilkinson, Bull.\ Am.\ Astron.\ Soc.\ 8, 351 (1976). 

\bibitem{smo77} G.F.~Smoot, M.V.~Gorenstein, and R.A.~Muller, Phys.\ Rev.\ Lett.\ 39, 898 (1977). 




\bibitem{kog93} 
A.~Kogut et al., Astrophys.\ J.\ 419, 1 (1993). 

\bibitem{ben03}
C.L.~Bennett et al., Astrophys.\ J.\ Suppl.\ 148, 1 (2003).

\bibitem{bra68}
T.N.~Bracewell and E.K.~Conklin, Nature 219, 1343 (1968).

\bibitem{hen68}
G.R.~Henry, R.B.~Feduniak, J.E.~Silver, and M.A.~Peterson, Phys.\ Rev.\ 176, 1451 (1968).

\bibitem{pee68}
P.J.E.~Peebles and D.T.~Wilkinson, Phys.\ Rev.\ 174, 2168 (1968).

\bibitem{wei72}
S.~Weinberg, {\it Gravitation and Cosmology} (Wiley, New York, 1972), pp.\ 520--522.

\bibitem{cha01}
A.~Challinor and F.~van Leeuwen, Phys.\ Rev.\ D 65, 103001 (2002).

\bibitem{bur06}
S.~Burles and S.~Rappaport, Astrophys.\ J.\ Lett.\ 641, L1 (2006).


\bibitem{mck79}
J.M.~McKinley, Am.\ J.\ Phys.\ 47, 602 (1979).

\bibitem{WMAP5}
E.~Komatsu et al., Astrophys.\ J.\ Suppl.\ 180, 330 (2009).

\bibitem{kno95} L.~Knox, Phys.\ Rev.\ D 52, 4307 (1995). 

\bibitem{cop07} C.J.~Copi, D.~Huterer, D.J.~Schwarz, and G.D.~Starkman,
Phys.\ Rev.\ D 75, 023507 (2007).

\bibitem{hin96} G.~Hinshaw et al., Astrophys.\ J.\ Lett.\ 464, 25 (1996).

\bibitem{tur91} M.S.~Turner, Phys.\ Rev.\ D 44, 3737 (1991). 

\bibitem{gri92} L.P.~Grishchuk, Phys.\ Rev.\ D 45, 4717 (1992).

\bibitem{lan96} D.~Langlois and T.~Piran, Phys.\ Rev.\ D 53, 2908 (1996). 

\bibitem{eri08} A.L.~Erickcek, M.~Kamionkowski, and S.M.~Carroll, Phys.\ Rev.\ D 78, 123520 (2008).

\bibitem{mer09} L.~Mersini-Houghton and R.~Holman, JCAP 02, 006 (2009). 

\bibitem{kas10} A.~Kashlinsky et al., Astrophys.\ J.\ Lett.\ 712, L81 (2010); F.~Atrio-Barandela et al.,
Astrophys.\ J.\ in press (2010);
A.~Kashlinsky, F.~Atrio-Barandela, D.~Kocevski, and H.~Ebeling, Astrophys.\ J.\ 691, 1479 (2009)
and  Astrophys.\ J.\ Lett.\ 686, L49 (2008).

\bibitem{wat09} R.~Watkins, H.A.~Feldman, and M.J.~Hudson, Mon.\ Not.\ R.\ Ast.\ Soc.\ 392, 743 (2009).



\end{thebibliography}
\end{document}